# Pseudo-spin Kondo effect versus hybridized molecular states in parallel Double Quantum Dots

Alexander W. Holleitner

Center for NanoScience (CeNS), LMU Munich, Geschwister-Scholl-Platz 1, 80539

Munich, Germany

Present: Department of Physics and California NanoSystems Institute, University of

California, Santa Barbara, CA 93106, USA

Alexander Chudnovskiy and Daniela Pfannkuche

I. Institut für Theoretische Physik, University Hamburg, 20355 Hamburg, Germany

Karl Eberl

Max-Planck-Institut für Festkörperforschung, 70569 Stuttgart, Germany

Robert H. Blick

Electrical and Computer Engineering, University of Wisconsin-Madison, Madison, WI

53706-1691, USA

A two quantum-dot device is coupled in parallel for studying the competition between the pseudo-spin Kondo effect and strongly hybridized molecular states. Cryogenic measurements are performed in the regime of weak coupling of the two dots to lead states under linear transport conditions. Detailed simulations verify the finding of the transition between the two different regimes.

## Introduction

One of the important problems in condensed matter physics is the Kondo problem [1] [2] elucidating the interplay between localized quantum states and delocalized mediating wave functions [3]. Many aspects of the physics involved can by now be tested using quantum dots which act as Kondo impurities at very low temperatures. [4] [5] [6] [7] Correlations between quantum states of delocalized lead electrons and localized electrons of a single quantum dot generate a resonant second order tunneling of electrons that screens the local spin of the dot. Experimentally this results in the linear conductance reaching the unitary limit $G = 2e^2/h$ at lowest temperatures. In the non-linear regime a zero-bias peak of the differential conductance with a width given by the Kondo temperature $T_K$ emerges. [5] [6] Expanding the dot system by defining two coupled quantum dots the orbital structure of the wave functions acquires spin-like features which can be expressed in terms of a pseudo-spin. [8] Previous theory work revealed the existence of this SU(4) pseudo-spin Kondo effect if the Coulomb interaction between the dots is large enough. [8] This Letter focuses on such a coupled double quantum dot which is connected to source and drain in parallel. The conductance through the device increases resonantly if the ground state energy does not change by the transfer of an electron between the two quantum dots. [9] As a consequence of the orbital Kondo correlations these two degenerate states play the role of pseudo-spin up and pseudo-spin down states. Fitting the theory of pseudo-spin to data in that regime shows that second order tunneling dominates the transport through the double quantum dot. Generally speaking, the interdot tunneling $t$ breaks the pseudo-spin Kondo correlations in the same way an external magnetic field does for the spin Kondo effect. At very large tunneling the states with an electron on either quantum dot are strongly hybridized and form a molecular ground state. [10] In that state, the pseudo-spin is frozen in the symmetrical combination of pseudo-spin up and pseudo-spin down direction, the orbital Kondo effect and, respectively, the second order tunneling through the device is suppressed. Again the experiment shows this transition in unambiguous agreement with the model.

**Experiment**

Two quantum dots are defined in a two-dimensional electron system (2DES) 90 nm below the surface of an AlGaAs/GaAs heterostructure via the biasing of Schottky-gates patterned on top of the semiconductor. Sheet density and mobility of the 2DES are found to be $\mu = 80 \, m^2/Vs$ and $n_S = 1.7 \times 10^{15} \, m^{-2}$ at $4.2 \, K$. As depicted in Fig. 1(a) and explicitly described in [11] [12] [13] the two dots are built in an Aharonov-Bohm geometry with tunable interdot tunneling $t$ by a two-step electron-beam writing process. In this geometry it is possible to measure correlated second order tunnel events (co-tunneling) not only for dot levels aligned to the Fermi levels $\mu_S$ and $\mu_D$ of source and drain (see Fig. 1(b)) but also for the off-resonant condition sketched in Fig. 1(c). In order to characterize the coupled quantum dot the conductance is measured from source to drain while sweeping the chemical potentials of the two dots separately. [13] [14] [15] [16] Latter is achieved by two electrostatically coupled top gates, i.e. applying two voltages $V_1$ and $V_2$ for $dot_1$ and $dot_2$, respectively. For coupled double dots this leads to phase diagrams with hexagonal patterns as depicted in Fig. 2(a). In the diagram each line indicates the phase boundary between two ground states of the double quantum dot. For lines parallel to (1) [(2)] the chemical potential of $dot_1$ [$dot_2$] is aligned to source and drain.

Consequently, single electrons are tunneling via this dot and the corresponding electron number is reduced by one, e.g. from $n$ to $n-1$ [$m$ to $m-1$] as well described by the orthodox theory[17]. A very distinct feature in this scenario is given by line (3) which refers to Fig. 1(c). These positions in the phase diagram enable highly correlated tunneling processes in which two electrons resonantly tunnel through the parallel two-dot system [9] and allow directly detecting the transition from delocalized pseudo-spin Kondo states to a localized ground state with a frozen pseudo-spin.

## Measurements

In order to verify this transition transport spectroscopy at cryogenic temperatures is performed. The two measured quantum dots reveal charging energies of $E_1 = e^2/C_1 \sim 2.22 \, meV$ ($C_1$ describes the total capacitance of $dot_1$) and $E_2 \sim 2.76 \, meV$ and single level spacings of about $\varepsilon \sim 240 \, \mu eV$ in a geometry analogous to Fig. 1(a). The base temperature of the $^3He/^4He$ dilution refrigerator is $T_{BATH} = 55 \, mK$ while the lock-in excitation voltage is $V_{AC} = 10 \, \mu V$ at $17 \, Hz$. The corresponding gray scale plot in Fig. 2(b) depicts conductance measurements in reference to the two gate voltages $V_1$ and $V_2$ where only tunneling processes along line (3) are resonant. The maximum conductance detected is of the order of $G = 0.1 \, e^2/h$ which is significantly below the maximum conductance quantum and is explained by the model below. But since the conductance along lines (1) and (2) is below the noise level of about $\Delta G \sim 0.002 \, e^2/h$ it already follows that first order tunneling through a single dot individually is suppressed.

## Theory

The four states of the double-dot system involved in the transport along line (3) are $(n, m)$, $(n-1, m)$, $(n, m-1)$ and $(n-1, m-1)$ as indicated in Fig. 2(a). In absence of the interdot tunneling the ground state energies of two configurations $(n-1, m)$ and $(n, m-1)$ are equal along the line. The interdot tunneling $t$ leads to the splitting of the otherwise degenerate energies. Instead of $E(n, m-1) = E(n-1, m)$ the mixed states denote as pseudo-spin up $|\uparrow\rangle$ and pseudo-spin down $|\downarrow\rangle$ with energies $\varepsilon_{\uparrow,\downarrow} = E(n, m-1) \mp t$, respectively. These two states constitute the low energy sector of the model. The high sector is given by the states $(n-1, m-1)$ and $(n, m)$. Since in the middle of line (3) the energy $E(n-1, m-1)$ is high with respect to the chemical potentials in the leads, $(n-1, m-1)$ is associated with an empty state. Then the Hamiltonian of the isolated double-dot system reads (1)

$$H_{dot} = \sum_{\sigma=\uparrow,\downarrow}(E(n,m-1) - E(n-1,m-1))\hat{n}_\sigma + [E(n,m) - E(n,m-1)]\hat{n}_\uparrow\hat{n}_\downarrow + t(\hat{n}_\uparrow - \hat{n}_\downarrow),$$

with $\hat{n}_\sigma = \hat{c}_\sigma^+\hat{c}_\sigma$ the fermion creation operators $\hat{c}_\uparrow^+ = |\uparrow\rangle\langle n-1, m-1| + |n, m\rangle\langle\downarrow|$ and $\hat{c}_\downarrow^+ = |\downarrow\rangle\langle n-1, m-1| + |n, m\rangle\langle\uparrow|$. Spin-Kondo correlations are not considered here. The tunneling between the leads and the double-dot is described by the Hamiltonian $H_{tun} = \Sigma_k \Sigma_{\nu=r,l} \Sigma_\sigma \{T_{k\sigma}^\nu \hat{a}_{\nu k}^+ \hat{c}_k + h.c.\}$, where $\nu = r, l$ relates to the right and the left lead respectively. $\hat{a}_{\nu k}^+$ characterizes the creation operator for an electron in the lead $\nu$ with the wave vector $k$. Note that the spin index of electrons in reservoirs is suppressed and that

the tunneling is assumed to be spin-independent. For a given mode $(v,k)$, however, the tunneling to the upper and lower pseudo-spin state differs. The total Hamiltonian $H = H_{dot} + H_{tun}$ is similar to the Anderson impurity model with Zeemann field $t = 1/2(\varepsilon_\uparrow - \varepsilon_\downarrow)$. Performing the Schrieffer-Wolff transformation the effective pseudo-spin Kondo Hamiltonian reads (2) $H_{pK} = \sum_{vv'} \sum_{kk'} \sum_{\mu=x,y,z} J^\mu_{vv'}(k,k') \hat{s}_\mu \hat{a}^+_{v',k'} \hat{a}_{v,k} + t \cdot \hat{s}_z$. The asymmetry of the tunneling of each mode in the reservoir to different dots leads to the wave vector dependence of the pseudo-spin Kondo constants. In the case of the highest possible asymmetry the modes of the reservoirs can be assigned with a pseudo-spin (orbital) index that remains conserved by tunneling. In that case the pseudo-spin Kondo effect is most pronounced. Even in the absence of the complete asymmetry, however, the pseudo-spin Kondo couplings do not vanish, thus implying the existence of pseudo-spin Kondo correlations. The pseudo-spin Kondo constants vanish only for the completely symmetric tunneling of each mode to both dots.

In order to describe the experimental data in Fig. 2(b), the two-particle co-tunneling contribution to the current through the double dot is calculated which is a precursor of the pseudo-spin Kondo effect, using the pseudo-spin Kondo Hamiltonian (2).[8] The co-tunneling contribution to the current follows as (3)

$$I_{cot} \propto \frac{F_B(\varepsilon_1) + F_B(\varepsilon_2)}{[\log(\max(T,2t)/T_K)]^2} [eV \frac{\sinh(2\beta t)\tanh(\beta t)(\cosh(\beta eV)-1)}{(\cosh(2\beta t)-1)(\cosh(\beta eV)-\cosh(2\beta t))} +$$

$$2t\tanh(\beta t) \frac{\cosh(\beta eV) - \exp(-\beta eV)}{\cosh(2\beta t) - \cosh(\beta eV)}],$$

where $T_K$ denotes the pseudo-spin Kondo temperature, $\beta = k_B T$ the thermal energy with $k_B$ the Boltzmann constant and $F_B(\varepsilon)$ the Boltzmann distribution insuring that the state with energy $\varepsilon$ is occupied. Without the interdot tunneling $t = 0$ the co-tunneling current is given by $I_{cot} \propto V/[\log(T/T_K)]^2$, whereas the current gets exponentially suppressed with interdot tunneling at $2\beta t > 1$, $I_{cot} \propto V/[\log(\max(T,t)/T_K)]^2 2\beta t e^{-2\beta t}$. The conductance of the double-dot between the two triple points is given by the sum of the contribution from the sequential tunneling $\sigma_{seq}$ and the co-tunneling contribution $\sigma_{cot} = dI_{cot}/dV|_{V=0}$ that is obtained by differentiating (3). Now the expression (4) $\sigma = \sigma_{seq} + \sigma_{cot}$ models the experimental data in Fig. 2(b) along line (3). The fit parameters are: the sequential tunneling rate $\Gamma_S$, the co-tunneling rate $\Gamma_{cot}$, the ground state energies of the isolated dots $E(n-1, m-1)$, $E(n, m-1) = E(n-1, m)$, $E(n, m)$, the interdot tunneling amplitude $t$ that simultaneously plays the role of Zeemann field for the pseudo-spin and the temperature $T$. The inset in Fig. 2(b) shows the comparison between the data along line (3) and the fit with dot energies which are equivalent to the experimentally extracted ones. Confirming that in this regime the conductance is governed by co-tunneling the fit yields tunneling rates of $\Gamma_1 < 0.013\ GHz$ and $\Gamma_{cot} = (3.33 \pm 0.02)\ GHz$. For considering only co-tunneling by definition, the fit stabilizes at a temperature reduced by about 40% which coincides with the difference between the bath temperature $T_{BATH} = 55 mK$ and the electron temperature in the leads $T = (95 \pm 20) mK$ [a] (see also (9)).

---

[a] Only filters at room temperature are integrated in the set-up.

Fig. 3 demonstrates the effect of increasing the interdot coupling. [11] In Fig. 3(a) the two dots exhibit strong pseudo-spin Kondo correlations (along line (3)) but also sequential tunneling processes contribute to the transport as indicated by the black arrow. Therefore, this measurement illustrates the transition between the pseudo-spin Kondo state and fully molecular modes inside the double quantum dot. Finally in Fig. 3(b) all boundaries of a two dot phase diagram are observable. Very distinct to the pseudo-Kondo regime in Fig. 2(b), also triple points appear as marked by white circles. At these points the conductance is maximum, since electrons from source can sequentially tunnel via three different double dot ground states to drain. [16] The top trace of Fig. 3(c) highlights a line plot from the phase diagram in Fig. 3(b) along line (3) from one triple point to another. Clearly the overall increase in conductance at these triple points is seen (indicated by white circles). In between both first and second order tunneling processes via states $(n, m-1)$ and $(n-1, m)$ give rise to conductance as expected for parallel quantum dots. [13] More precisely, the theory fit (straight line) which follows nicely the experimental data (open boxes) reveals that first order tunneling events dominate this regime with $\Gamma_1 = (0.42 \pm 0.01)\, GHz$ and $\Gamma_{cot} < 1\, MHz$. Reducing the interdot coupling systematically leads to the lowest trace which corresponds again to Fig. 2(b).

Expression (4) and therefore the model successfully fits all curves in Fig. 3(c) by a monotonic increase of the interdot coupling $t$ from the lowest trace to the top most trace while all other fitting parameters stay within the experimental error. More explicitly, the fit states that the co-tunneling contribution to the conductance through the parallel double quantum dot is inversely determined by the magnitude of the interdot tunneling. As already mentioned above, in Fig. 2(b) the sequential tunneling is significantly suppressed whereas the co-tunneling contribution is pronounced. An asymmetry of the tunnel couplings between each dot and the right and the left lead explains this imbalance which is essential to the whole experiment. Assuming $T_{1l}/T_{1r} = T_{2r}/T_{2l} = \lambda > 1$, where $r,l$ denotes the right and the left lead and 1, 2 the dot numbers [b] Then the sequential tunneling contribution, being defined by the square of the smaller tunnel coupling, is proportional to $|T_{1r}|^2 + |T_{2l}|^2$ whereas the co-tunneling contribution is dominated by the product of the largest tunneling couplings $|T_{1l}|^2|T_{2r}|^2$. Therefore the ratio of resonant contributions of the co-tunneling and the sequential tunneling is $\lambda^4/2$ in that case.

## Summary

In summary this work covers experiment and theory on pseudo-spin Kondo correlations in a semiconductor double quantum dot. By altering the interdot and dot-lead coupling cryogenic transport spectroscopy traces the transition from a pseudo-spin Kondo ground state to a ground state with frozen pseudo-spin and suppressed pseudo-spin Kondo correlations. A detailed theoretical analysis confirms the finding of this transition.

We like to thank J. P. Kotthaus and D. D. Awschalom for continuous support and stimulating discussions. We acknowledged financial support by the Deutsche

---

[b] This assumption is supported by non-linear transport measurements which are not shown here.

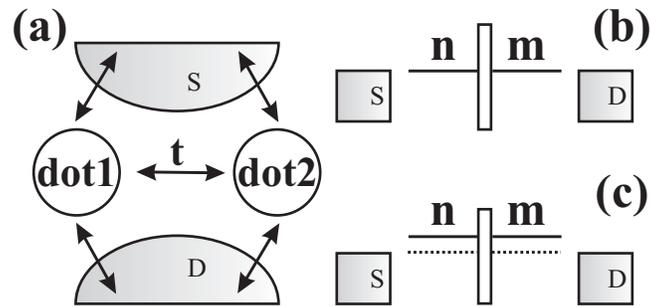

**Fig. 1: (a) Parallel double quantum dot with each dot connected to the chemical potentials $\mu_S$, $\mu_D$ of source and drain. The interdot tunnel coupling is denoted by $t$. (b) Resonant first order level scheme: quantum level $n$ ($m$) with $n$ electrons on $dot_1$ ($dot_2$) is aligned to the chemical potentials of the leads. The open box sketches the interdot tunneling barrier (c) For levels being off-resonant the resonant state generated by correlated tunneling processes at the Fermi level can be detected. Hereby the phase transition between first and second order tunneling is traced.**

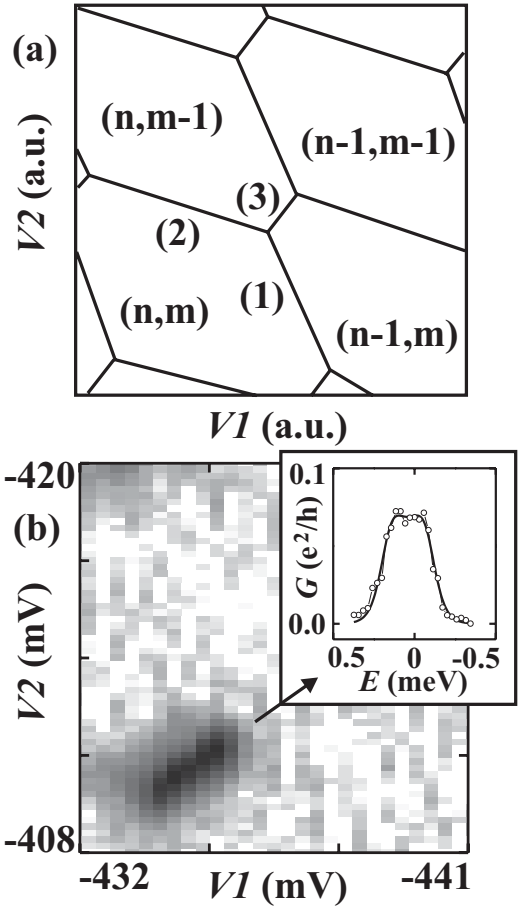

**Fig. 2: (a) Schematic phase diagram of a double quantum dot: black lines illustrate phase boundaries (1), (2) and (3) between two ground state configurations ($n$, $m$), respectively. (b) Pseudo-Kondo regime: logarithmic gray scale plot of a measured phase diagram (white $\leq 0 \leq 0.1\, e^2/h \leq$ black). Only second order tunneling processes occur along the degenerate phase boundary between dot configurations ($n, m-1$) and ($n-1, m$). Inset depicts the comparison of the model (straight line) to experimental data (open circles) along line (3).**

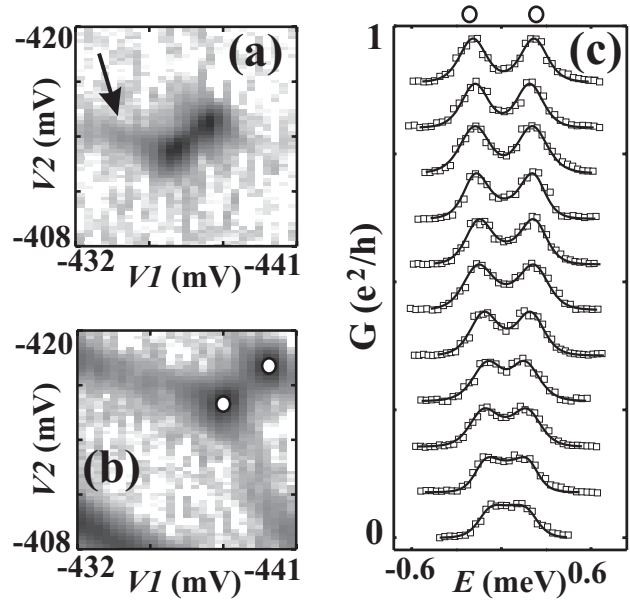

**Fig. 3: (a) Phase transition: in addition to second order processes also first order processes (e.g. black arrow) are detectable. (b) State with frozen pseudo-spin: along all phase boundaries only first order processes can be traced. White circles denote triple points where three charge configurations are degenerate. (c) Single conductance traces (data: open boxes, theory: straight lines): topmost trace corresponds to Fig. 3 (b) and lowest to Fig. 2 (b), showing the transition from the molecular ordered state to the pseudo-Kondo regime. Line plots are displayed with a small offset for clarity.**